\documentclass[prl,preprint,superscriptaddress]{revtex4}
\usepackage{graphicx}
\usepackage{dcolumn}
\usepackage{amssymb}
\usepackage{bm}
\usepackage{pifont}
\usepackage{epsfig}
\usepackage{psfrag}
\usepackage{epstopdf}
\usepackage{wasysym}
\usepackage[usenames]{color}
\begin{document}

\title{Site-Specific Colloidal Crystal Nucleation by Template-enhanced Particle Transport}

\author{Chandan K Mishra} \thanks{To whom correspondence should be addressed} \email{mishrachandan23@gmail.com}
\affiliation{Chemistry and Physics of Materials Unit, Jawaharlal Nehru Centre for Advanced Scientific Research, Jakkur, Bangalore - 560064, INDIA}

\author{A K Sood}
\affiliation{Department of Physics, Indian Institute of Science, Bangalore - 560012}
\affiliation{International Centre for Materials Science, Jawaharlal Nehru Centre for Advanced Scientific Research, Jakkur, Bangalore - 560064, INDIA}

\author{Rajesh Ganapathy} \thanks{To whom correspondence should be addressed} \email{rajesh.ganapathy@gmail.com}
\affiliation{International Centre for Materials Science, Jawaharlal Nehru Centre for Advanced Scientific Research, Jakkur, Bangalore - 560064, INDIA}
\affiliation{Sheikh Saqr Laboratory, Jawaharlal Nehru Centre for Advanced Scientific Research, Jakkur, Bangalore - 560064, INDIA}

\date{\today}

\draft

\begin{abstract}
The monomer surface mobility is the single most important parameter that decides the nucleation density and morphology of islands during thin film growth. During template-assisted surface growth in particular, low surface mobilities can prevent monomers from reaching target sites and this results in a partial to complete loss of nucleation control. While in atomic systems a broad range of surface mobilities can be readily accessed, for colloids, owing to their large size, this window is substantially narrow and therefore imposes severe restrictions in extending template-assisted growth techniques to steer their self-assembly. Here, we circumvented this fundamental limitation by designing templates with spatially varying feature sizes, in this case moir\'e patterns, which in the presence of short-range depletion attraction presented surface energy gradients for the diffusing colloids. The templates serve a dual purpose, first, directing the particles to target sites by enhancing their surface mean free paths and second, dictating the size and symmetry of the growing crystallites. Using optical microscopy, we directly followed the nucleation and growth kinetics of colloidal islands on these surfaces at the single-particle level. We demonstrate nucleation control, with high fidelity, in a regime that has remained unaccessed in theoretical, numerical and experimental studies on atoms and molecules as well. Our findings pave the way for fabricating non-trivial surface architectures composed of complex colloids and nanoparticles.
\end{abstract}

\maketitle

Realizing ordered surface nano- and microstructures of well-defined size and shape from the autonomous assembly of their constituent building blocks continues to remain one of the key challenges in materials science {\cite{lagally_science_1997, kern_nature_2005, dieter_RMP_1999, brune_surface_science_1998}. With regard to atomic/molecular surface assembly, using templates that preferentially enhance crystal nucleation events at specific sites is a proven approach towards realizing mesoscopically ordered structures like quantum dot arrays and supramolecular nanoassemblies \cite{brune_nature_1998, whitesides_nature_1999}. Extending this technique to tailor surface structures composed of nanoparticles and colloids is highly desirable for applications that include sensors, structural color-based filters and optoelectronic devices \cite{blaaderen_nature_1997, yodh_PRL_2000, GiRaYi_AsiaMat_2011}. Although recent experiments find that colloidal and atomic thin film growth on homogeneous surfaces obeys identical scaling laws \cite{ganapathy_science_2010, itai_PNAS_2013}, translating concepts gleaned from site-specific nucleation studies on atoms/molecules to colloids is anything but easy. To achieve site-specific nucleation with high fidelity, the particles' surface mean free path, $L$, should be larger than the distance between the preferential nucleating sites $L_p$. In conventional surface growth studies, $L$ is set by the ratio of the monomer surface diffusion constant to its deposition flux $D/F$ with $L \sim \left({D\over F}\right)^{1/6}$ for $D/F > 10^4$ \cite{brune_surface_science_1998, stanley_PRB_1994}. For atoms, a broad window of $D/F$ values spanning $10^{-1}-10^9$ can be readily accessed and thus, $L$ can be easily tuned to be larger than $L_p$. However for colloids, owing to their large size, $D$ is small and the $D/F$ window is limited to $10^{-1}-10^4$. To further complicate matters, over this window, simulations, atomic and colloid experiments find deviations from mean-field scalings and $L$, or equivalently the island density, is found to saturate with decreasing $D/F$ \cite{brune_surface_science_1998, ganapathy_science_2010}. In fact for micrometer-sized colloids, the maximum $L$  is $\sim5-7$ particle diameters only. Consequently, it is as yet unclear whether site-specific nucleation is even a viable strategy to fabricate mesoscopically-organized  structures made of nano- and microscale particles.

A plausible route to help alleviate the restrictions on $L$, imposed by $D/F$, is to utilize surfaces with energy gradients to transport particles to desired locations. Using the facile replica imprinting technique \cite{whitesides_science_1996}, we present a new design principle based on templates with spatially varying feature sizes which in the presence of short-range depletion attraction induced activation energy gradients for the diffusing colloids. Our substrates comprised of linear and square moir\'e patterns on polymethylmethacrylate (PMMA) layers spin-coated on glass coverslips. We fabricated these patterns, by first transferring a linear (square) array of trenches (holes), with periodicity $\lambda$, from a master grating to the PMMA substrate \cite{yodh_PRL_2000}. This was followed by a second imprint at an angle, $\theta$, relative to the first, which resulted in a long wavelength modulation of the trench (hole) depths (see Materials and Methods). While $\lambda$ decides the symmetry of the growing crystallites, the wavelength of the moir\'e pattern sets $L_p$, and can be independently controlled by varying $\theta$ as seen in Fig. \ref{Figure1}a and b. Next, we sedimented colloidal particles (silica, diameter $\sigma = 940$ nm, \cite{blaaderen_langmuir_2009}) in the presence of a depletant polymer (Sodium carboxylmethyl cellulose, $R_g = 60$ nm) on these substrates. Particle diffusion on templated surfaces has an activated form, $D= D_0 e^{-E_a/{k_BT}}$ where $E_a$ is the barrier height, $D_0$ is the attempt frequency and $k_BT$ is the thermal energy \cite{brune_surface_science_1998}. Apart from the polymer concentration $c$, the strength of the depletion attraction between the colloids and the substrate is also proportional to the excluded overlap volume that is freed up when they come in contact \cite{asakura_JCP_1954}. Thus, on moir\'e patterns, gradients in trench (hole) depths induced gradients in $E_a$ for the diffusing colloids and results in particle migration to the nearest energy minima (Fig. \ref{Figure1}c). 

We confirmed the enhancement in particle mean free paths by analysing the motion of individual particles on a linear moir\'e substrate, with $L_p = 42\sigma$ (Fig. \ref{Figure1}d and see Supplemental). The grey shaded region in Fig. \ref{Figure1}d represents regions of the template with a high $E_a$. While the particles that land within this region stay localised, those that land outside (regions of low $E_a$), migrate nearly $25\sigma$ to the high $E_a$ regions and are subsequently trapped. On a linear array of trenches (no surface energy gradients) and for the same $c$ and $\lambda$ as the moir\'e templates, we found $L\sim5-7\sigma$ (see Supplemental). Figure \ref{Figure1}e shows a snapshot of particles localised in trap sites on moir\'e templates. By quantifying the variation in $D$ across $L_p$, we estimated $\Delta E_a \sim 0.5 k_BT$ (see Supplemental).

The accumulation of particles in traps also increased the likelihood of crystal nucleation events at these sites. For site-specific nucleation to occur with a high fidelity, however, an optimal balance between $F$ and the particle surface mean free path, albeit enhanced here, is needed \cite{brune_surface_science_1998, stanley_PRB_1994}. Figure \ref{Figure2}a and b shows representative snapshots of crystals grown on linear moir\'e templates with $L_p = 42\sigma$ at identical $c$ and surface coverage, $\Theta$, and for two different $F$s, respectively. At the smaller $F$, particles were able to migrate to traps prior to encountering arriving monomer(s) and the crystals formed are compact and confined to trap sites (Fig. \ref{Figure2}a and see Supplemental). For the higher $F$, however, this is not the case and the crystals within the traps are more ramified and colloidal islands with six-fold symmetry (highlighted by circles in Fig. \ref{Figure2}b) nucleated outside the traps as well. We parametrized the fidelity of nucleation events at trap sites by the nucleation control efficiency, $x_{NCE} = N_{m}/N_{Tot}$. Here, $N_m$ is the subset of particles that migrated to traps in a time interval $\Delta t$ and then crystallised and $N_{Tot}$ is the total number of particles that landed outside traps over the same time interval \cite{heuer_PRL_2007, heuer_PRB_2009, x_NCE}.   

Next, we carried out thin film growth experiments on linear moir\'e templates for different $c$, $F$ and $L_p$ values. To succinctly capture the dependence of $x_{NCE}$ on these parameters and to isolate the contribution arising from the enhancement in mean free paths, we plotted $x_{NCE}$ versus $p^* = {L_p\over L}$. Here, $L$ is the mean free path of particles under identical experimental conditions but in the absence of energy gradients. $L$ was obtained by measuring the density of critical clusters, $n_c$, at the onset of coalescence (see Supplemental). An island with $i$-particles is termed a critical cluster if the addition of another particle makes it stable against disintegration. We found $i=2$ with and without surface energy gradients (see Supplemental). In Figure \ref{Figure2}c, we compare $x_{NCE}$ versus $p^*$ from our experiments with simulation and experimental results for site-specific nucleation and growth of organic molecules. The data points denoted by $1$ and $2$ correspond to Fig. \ref{Figure2}a and b, respectively. Strikingly, as opposed to previous studies where complete nucleation control ($x_{NCE} = 1$) \cite{heuer_PRL_2007} was possible only for $p^*\leq1$, on moir\'e patterns $x_{NCE} \approx 1$ until $p^*= 6$ and beyond which it decreases. Further, the lateral flux of particles to traps on moir\'e templates also effectively reduced the time taken to complete the nucleation process.

These observations clearly exemplify the efficacy of our approach in achieving site-specific nucleation for colloidal particles over a range of $L_p$ values. In order to shed light on the nucleation kinetics, we mimicked atomic heteroepitaxy experiments \cite{kern_PRB_2000, rousset_crystal_growth_2005, rohart_physique_2005} and measured $n_c$ as a function of $c$, at constant $F$, on square moir\'e patterns with $L_p = 16\sigma$. Here, $c$ plays the role of an inverse temperature. At the largest $c$s studied ($c \geq 0.27$ mg/ml), particles are unable to overcome the energy barriers for surface diffusion \textit{even} in the low $E_a$ regions and \ the moir\'e template thus acts like a homogeneous surface. On such surfaces, for small $D$, atomic and colloid experiments follow rate equation predictions and find that $n_c$ depends only on $\Theta$ and saturates to a constant, $n_c \approx 0.03$ for $i=1$ \cite{brune_surface_science_1998, rousset_crystal_growth_2005}. This was indeed found to hold in our experiments as well (triangles in Fig. \ref{Figure3}a). For $0.2 < c < 0.27$ mg/ml, although particles diffuse, their mean free path remains smaller than $L_p$ and nucleation events continue to occur at random locations (panel labeled 3 in Fig. \ref{Figure3}a). Akin to the behaviour on a homogeneous surface $n_c$ versus $c$ exhibits an Arrhenius-like dependence. We found a second plateau in $n_c$ for $c$ lying between $0.1-0.2$ mg/ml (green squares and red circle in Fig. \ref{Figure3}a). In this regime, particles perceive the heterogeneous nature of the surface energy landscape and thus their diffusivities are preferentially smaller at trap sites, which subsequently resulted in site-specific nucleation (panel 2 in Fig. \ref{Figure3}a). Finally, for the lowest $c$s studied ($c\leq0.1$ mg/ml), particles are oblivious to the underlying surface topography and $n_c$ decreases owing to an increase in the critical cluster size ($i\geq4$) (panel marked 1 in Fig. \ref{Figure3}a).    

Although the overall shape of the nucleation curve (red line in Fig. \ref{Figure3}a) for colloids is strikingly similar to that seen for atoms, there are fundamental differences. The black line in the schematic in Fig. \ref{Figure3}b shows $n_c$ versus $D/F$, or the scaled $1/T$, for homogeneous surface growth. When dimers are stable and also immobile, $L \propto D/F^{1/6}$ for $D/F>10^4$ \cite{brune_surface_science_1998}. In site-specific nucleation studies performed hitherto, since $L$ is decided by $D/F$ alone, a plateau in $n_c$ is observed when $L\sim L_p$ (green line in Fig. \ref{Figure3}b) and more importantly this plateau is restricted to the yellow shaded region, corresponding to $L\geq L_p$, of Fig. \ref{Figure3}b. On moir\'e templates, however, the enhancement in particle mean free path due to surface energy gradients, corresponds to a larger effective $D/F$ and the plateau in $n_c$ should therefore lie below the homogeneous nucleation curve (blue shaded region of Fig. \ref{Figure3}b). In line with expectations, $n_c$ predicted by homogeneous nucleation (magenta circles in Fig. \ref{Figure3}a) for $D/F$s corresponding to the plateau (blue shaded region) are indeed larger. 
                    
In the plateau regime of $n_c$, since crystalline islands are periodically spaced and have nearly identical monomer capture rates the island size distribution is expected to be narrower than on a homogeneous surface \cite{brune_surface_science_1998, rousset_crystal_growth_2005}. We measured the island size distribution on square moir\'e patterns with $L_p = 32\sigma$ by clustering particles based on their local bond-order parameter (Fig. \ref{Figure3}c and inset) and compared it with homogeneous surface growth experiments for the same $\Theta$ (see Supplemental) \cite{ganapathy_science_2010}. While we found a broad range of island sizes on homogeneous surfaces (black circles in Fig. \ref{Figure3}d) \cite{ganapathy_science_2010}, on moir\'e templates the distribution was peaked with a maximum that roughly coincided with the trap size (green circles in Fig. \ref{Figure3}d). Further, at the same $\Theta \sim 50\%$, we found islands on moir\'e templates to be more compact with a fractal dimension $d_f \sim 2$ while on the homogeneous surface we found $d_f \sim 1.7$ in agreement with theoretical predicitions (green and black circles in Fig. \ref{Figure3}e, respectively) \cite{stanley_PRB_1994}. With continued particle deposition, crystals with hexagonal order nucleated and grew outside the traps and subsequent layers were found to be in registry with the underlying symmetry (see Supplemental). We found this to be true even on templates with complex moir\'e periodicities. Figure \ref{Figure3}f shows a representative image of the third layer of crystals grown on a template that was fabricated by making multiple imprints at different $\theta$s. The symmetry and the width of the crystals reflect the underlying substrate periodicity with a motif: 8$\sigma\Box\rightarrow$8$\sigma\varhexagon\rightarrow$8$\sigma\Box\rightarrow$4$\sigma\varhexagon$.

Collectively, non-trivial substrate topographies, realized via a relatively simple approach, in the presence of short-range depletion interactions transported particles to desired locations and helped achieve site-specific nucleation with high fidelity \textit{even} for micron-sized colloidal particles. Depletion interactions being sensitive only to the local geometry \cite{pine_nature_2010, granick_nature_2011, pine_softmatter_2011}, a feature already exploited here, we believe our approach offers unparalleled opportunities in directing the self-assembly of complex colloids regardless of their surface chemistry and composition \cite{manoharan_science_2003, manoharan_science_2015}. The idea outlined here, however, is far more generic. By suitable manipulation of energy barriers for surface diffusion, control over nucleation density can be exercised over a substantially broader range of $D$ and $F$ values. It is, well-known that in atomic heteroepitaxy strain fields around misfit dislocations lead to directional adatom currents \cite{wolf_surface_science_1997, clemens_PRB_2004}. It is tempting to speculate if this can be exploited to guide growth for small $D/F$ values. In the context of nanoparticles, techniques for creating binding energy gradients by controlling the density of ligands on the surface already exists \cite{genzer_review_2012}. In light of our findings, we believe this approach should now be exploited in guiding the self-assembly of nanoparticles as well.  

\textbf{Acknowledgments} We thank Prof. Arjun G Yodh, Prof. Vinothan Manoharan and Prof. Prerna Sharma for useful discussions. C.K.M. thanks K. Hima Nagamanasa and Shreyas Gokhale for fruitful interactions. C.K.M. thanks CPMU, JNCASR for financial support.  A.K.S. thanks Department of Science and Technology, India for support under a J. C. Bose Fellowship and R.G. thanks SSL and ICMS, JNCASR for financial support.

\newpage
\begin{figure}[htbp]
\includegraphics[width=0.9\textwidth]{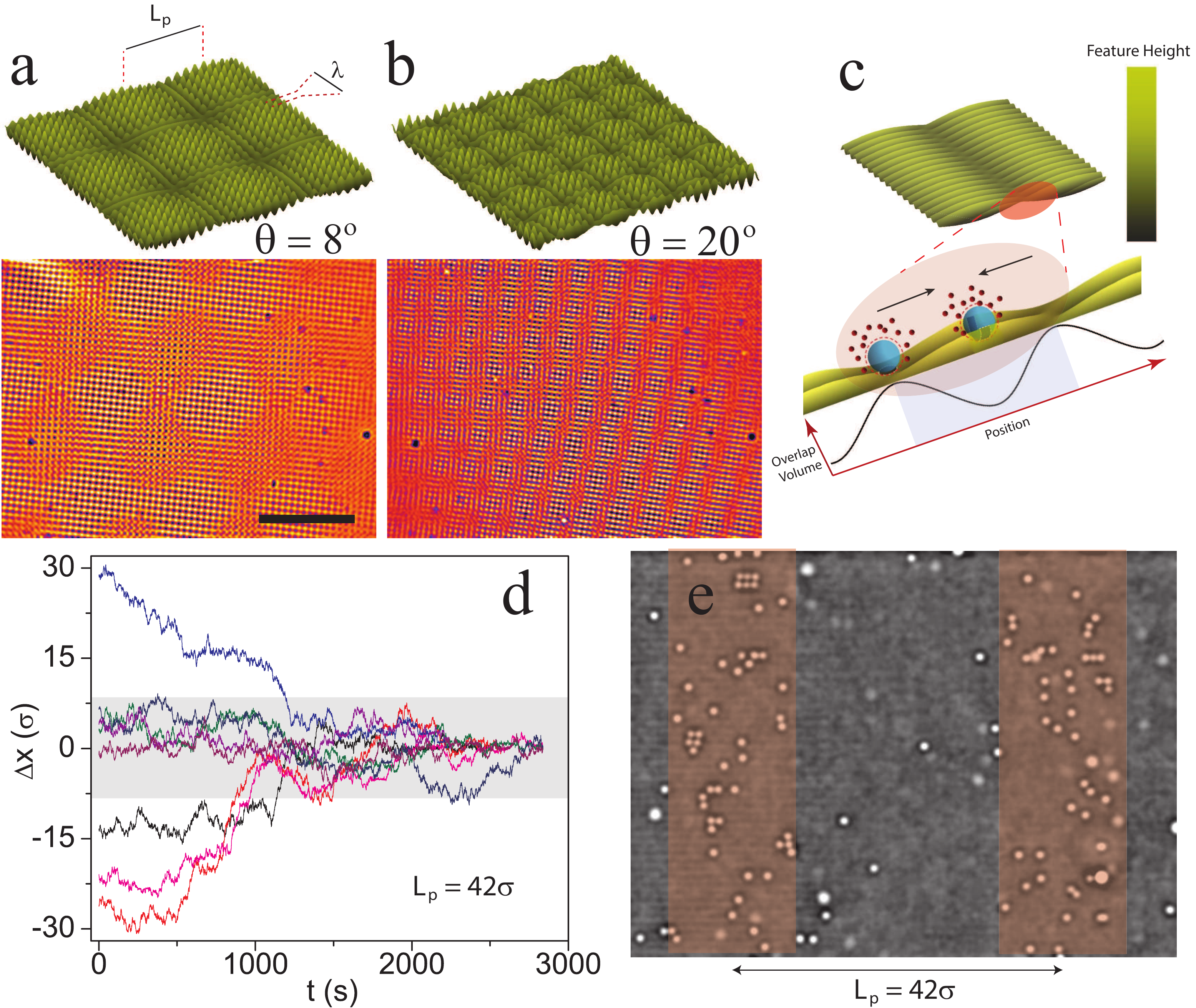}
\caption{\textbf{Particle locomotion to traps induced by surface energy gradients}.\textbf{ (a)} and \textbf{(b)} show representative optical micrographs of square moir\'e patterns (bottom panel) and the corresponding computer-generated topography maps (top panel) for two different $\theta$s. The scale bar in \textbf{(a)} represents 15 $\mu$m. \textbf{(c)} Schematic representing particle-substrate interactions on a linear moir\'e pattern. The overlap volume increases in the direction of the arrows and results in a net migration of particles to regions of high overlap (high $E_a$). \textbf{(d)} Particle trajectories on a linear moir\'e pattern. The grey region corresponds to regions of the pattern with high $E_a$.\textbf{ (e)} Representative image of particle localisation on moir\'e templates.}
\label{Figure1}
\end{figure}

\newpage
\begin{figure}[htbp]
\includegraphics[width=1\textwidth]{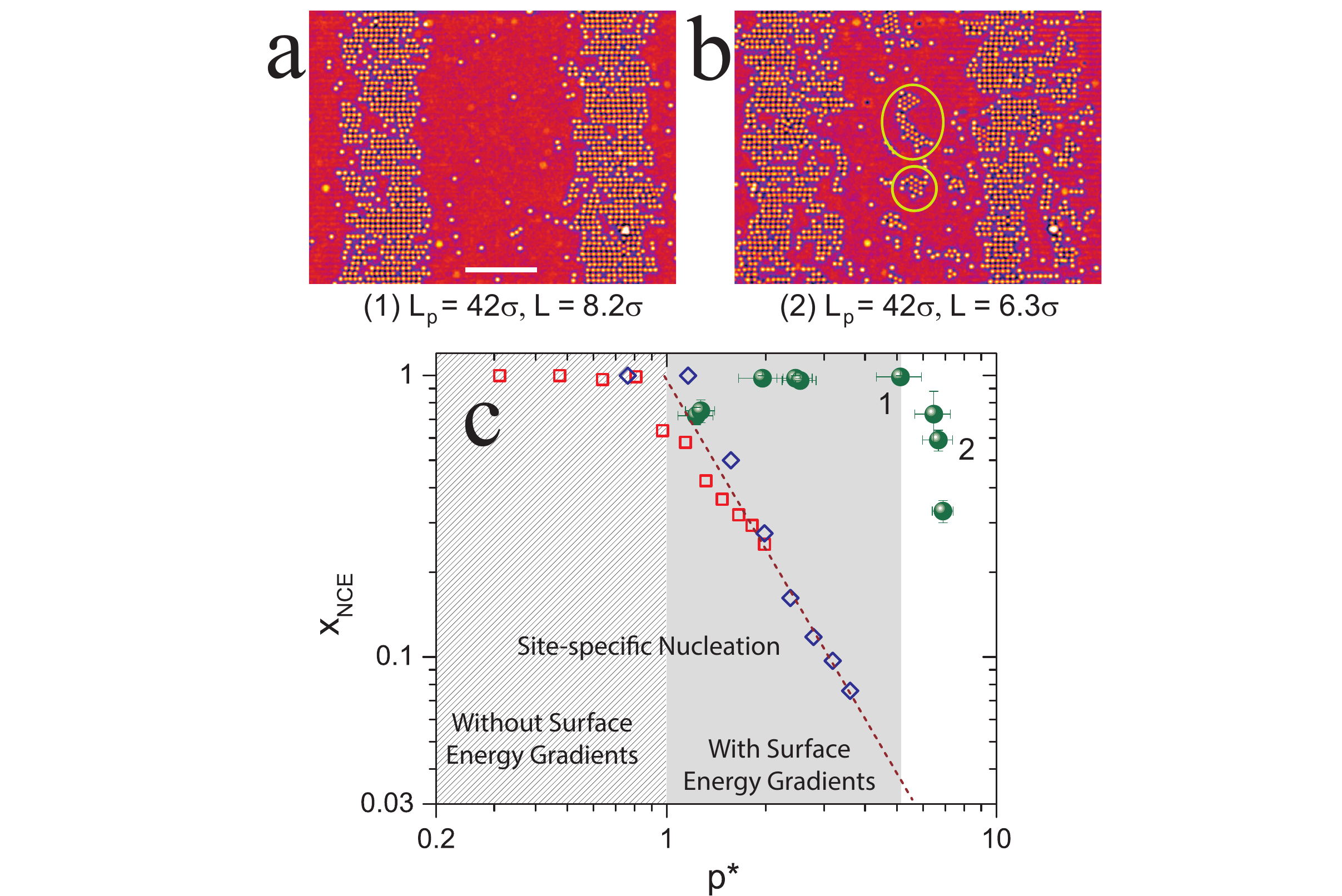}
\caption{\textbf{Nucleation control on moir\'e patterns.}\textbf{ (a)} and \textbf{(b)} Representative images of crystals growth on linear moir\'e patterns at constant $c$ and for two different $F$s. Since, $\lambda = \sigma$, the templates promoted the growth of crystals with square symmetry. (b) Loss of nucleation control results in the nucleation of hexagonally ordered crystallites outside of the traps and is highlighted by circles \textbf{(c)} Comparison of $x_{NCE}$ versus $p^*$ for conventional site-specific nucleation studies and on moir\'e patterns. The red and blue squares correspond to experiments and simulation results for vapor deposition of organic molecules (adapted from \cite{heuer_PRL_2007}). Here, nucleation control is lost ($x_{NCE}<1$) beyond the striped region. The green circles correspond to colloid experiments on linear moir\'e patterns for various $L_p$, $c$ and $F$ values. Owing to the enhancement in particle mean free path on these substrates, $x_{NCE} < 1$ only for $p^*>6$..}
\label{Figure2}
\end{figure}

\newpage
\begin{figure}[htbp]
\includegraphics[width=0.65\textwidth]{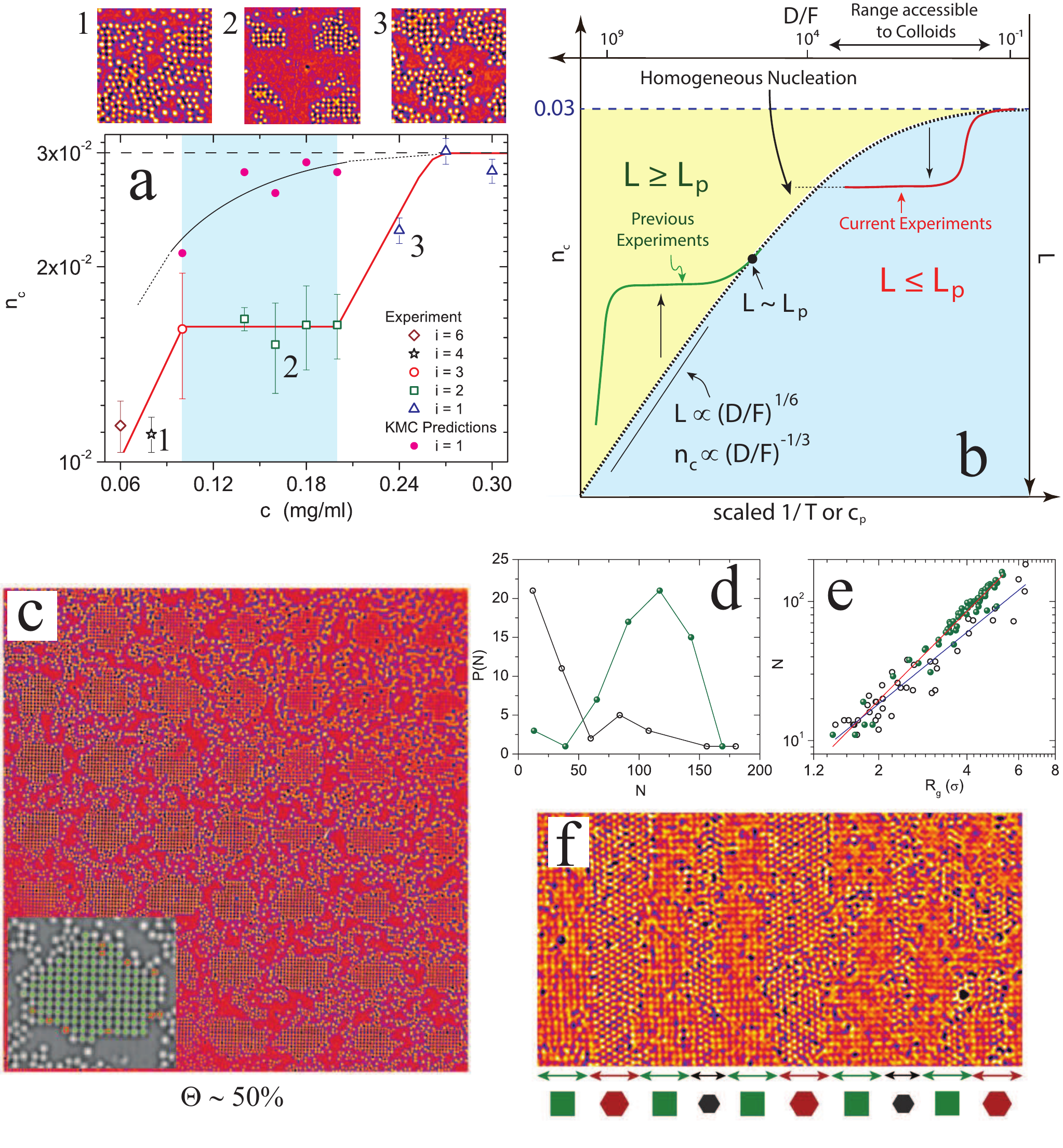}
\caption{\textbf{Nucleation and island growth on square moir\'e patterns.} \textbf{(a)} $n_c$ versus $c$ at fixed $F$. The legends in \textbf{(a)} represent the size of the critical cluster. The red line is a guide to the eye. The blue shaded region corresponds to the regime of organized growth. The magenta circles within the blue shaded region represent the expected $n_c$ from homogeneous nucleation, for $D/F$s in the plateau region (Adopted from \cite{brune_surface_science_1998}). \textbf{(b)} Schematic of nucleation curves for homogeneous nucleation (black curve) and heterogeneous nucleation with and without energy gradients shown by red and green curves, respectively. Mean-field scaling predictions for $L$ and $n_c$ for $D/F> 10^4$ is also shown. \textbf{(c)} Representative snapshot of island growth on square moir\'e patterns with $L_p = 32\sigma$. The inset shows particles clustered based on their bond-order parameter. Green represents particles with $\Psi _4>0.7$ and red represents particles with $\Psi_6>0.7$. \textbf{(d)} and \textbf{(e)} Island size distributions and fractal dimensions on square moir\'e patterns (green circles) and on hexagonal array of holes \cite{ganapathy_science_2010} (black circles) for $\Theta = 50\%$, respectively. \textbf{(f)} Colloidal crystal heterostructures grown on moir\'e templates. The image corresponds to the third layer of particles from the template. The size of the squares and hexagons represents the width of the strip. }  
\label{Figure3}
\end{figure}

\end{document}